\documentclass[a4paper]{panl}
\usepackage{cite}
\usepackage{wrapfig}
\usepackage{graphicx}
\usepackage{amssymb}
\usepackage{amsfonts}
\usepackage{amsmath}
\usepackage{longtable}
\usepackage{rotating}
\usepackage{lscape}
\usepackage{epsfig}
\usepackage{multirow}

\def\mz {M_{\scriptscriptstyle{Z}}}

\newcommand{\nll}{\nonumber \\}

\newcommand{\Ll}{k_1}
\newcommand{\Lll}{k_2}
\newcommand{\Llll}{k_3}
\newcommand{\Kl}{k_{11}}
\newcommand{\Kll}{k}

\newcommand{\sqrtLsmm}{\sqrt{\lambda}}
\newcommand{\cmi}{c^-}
\newcommand{\cpl}{c^+}

\originalTeX
\begin{document}
% Journal sections (see http://pkp.jinr.ru/index.php/PEPAN_LETTERS/about/editorialPolicies#focusAndScope)
\issuearea{Physics of Elementary Particles and Atomic Nuclei. Theory}
% or in Russian
%\issuearea{ФИЗИКА ЭЛЕМЕНТАРНЫХ ЧАСТИЦ И АТОМНОГО ЯДРА. ТЕОРИЯ}

\title{{\tt SANC}: the process $ \gamma \gamma \rightarrow ZZ $ \\
{\tt SANC}: процесс $ \gamma \gamma \rightarrow ZZ $}
\maketitle
\authors{
\fbox{D.\,Bardin$^{a}$},\footnote{E-mail: sanc@jinr.ru}, S.\,Bondarenko$^{b}$, 
\fbox{P.\,Christova$^{a}$}, L.\,Kalinovskaya$^{a}$, W.\,von Schlippe$^{d}$, E.\,Uglov$^{a,}$.}
\from{
$^{a}$ Dzhelepov Laboratory for Nuclear Problems, JINR,      \\
        ul. Joliot\d Curie 6, RU\d 141980 Dubna, Russia;}
\vspace{-3mm}
\from{$^{b}$ 
Bogoliubov Laboratory of  Theoretical Physics, JINR,  \\ 
        ul. Joliot\d Curie 6, RU\d 141980 Dubna, Russia;         \\
$^{c}$  Petersburg Nuclear Physics Institute, Gatchina, 188300, Russia.
}

\begin{abstract}
% Russian translation of the abstract
  Рассматривается внедрение процесса $\gamma\gamma \to ZZ$ в систему  {\tt
  SANC} на однопетлевом уровне точности в мультиканальном подходе. Полученные
  однопетлевые скалярные форм факторы могут быть использованы в любом канале
  после соответствующей перестановки их аргументов -- Мандельштамоских
  переменных $s,t,u$. Для проверки корректности результатов наблюдается
  независимость скалярных форм факторов от калибровочных параметров и
  выполнение тождества Варда (поперечность по внешнему фотону). Представлен
  полностью аналитический результат для тензорной структуры, ковариантной и
  спиральных амплитуд для данного процесса. Проведено расширенное сравнение
  аналитических и численных результатов с имеющимися данными в литературе. 
\vspace{0.2cm}

The implementation of the process  $\gamma\gamma \to ZZ$ at the one-loop level 
within  {\tt SANC} system multi-channel approach  is considered.
 The derived one-loop scalar form factors can be used for any cross channel
after an appropriate   permutation of their arguments -- Mandelstam variables $s,t,u$. 
 To check of the correctness of the results we observe the independence of the scalar 
form factors on the gauge parameters and the validity of Ward identity
(external photon transversality). 
We present the complete analytical results for the covariant and tensor structures 
and helicity amplitudes for this process.
We make an extensive comparison of our analytical and numerical results with
those existing  in the literature.
\end{abstract}
\vspace*{6pt}


%=========================================================================

\section{Introduction}
\label{sec:intro}

Physics with $\gamma\gamma$  collider
\cite{Ginzburg:1981ik},
\cite{Ginzburg:1981vm},\cite{Ginzburg:1982yr}
\cite{Telnov:2014hfa} and
with $e^{+} e^{-}$ linear collider 
%\cite{Accomando:1997wt},
\cite{Battaglia:2004mw},
\cite{MoortgatPick:2005cw},
\cite{Moortgat-Picka:2015yla}
always demonstrated the great interest
to establish the effects from transversal and longitudinal polarization.
We begin to create the theoretical support for the $\gamma\gamma$  colliders:
first of all it is {\tt SANC} modules  for processes $\gamma\gamma \to \gamma\gamma(\gamma Z ,ZZ$
at the one loop level \cite{Bardin:2009gq},\cite{Bardin:2012mc}.
Second step will be MC generator with these modules taking into account the polarization effects 
for $\gamma\gamma$ and $e^{+} e^{-}$ physics.

In this article we describe some results obtained with
{\tt SANC} ({\it Support of Analytic and Numerical calculations for
  experiments at Colliders}) ---
a network system for semi-automatic calculations for processes of
elementary particle
interactions at the one-loop precision level, see ~\cite{Andonov:2004hi}.
The corresponding  FORTRAN modules ~\cite{Andonov:2008ga}
 for processes at the one loop level can be downloaded by request.

 All calculations at the one-loop precision level are realized in the
 spirit of the  
book ~\cite{Bardin:1999ak} in the $R_\xi$ gauge and all the results are
reduced up  
to the scalar Passarino--Veltman functions: $A_0,\,B_0,\,C_0,\,D_0$
~\cite{Passarino:1978jh}.  
These two distinctive features allow one to perform several checks: e.g. to test
gauge invariance by observing the cancellation of gauge parameter dependence,  
to test symmetry properties and validity of the Ward identities,
all at the level of analytical expressions. 
The {\tt SANC} system is  based on  FORM ~\cite{Vermaseren:2000nd}  applications.
These applications had to be modularized as procedures in a most universal way 
so as to be used as building blocks for the computation of more complex
processes. 
We used a  covariant method for calculating helicity amplitudes (HA) presented in
\cite{Vega:1995cc}. 
The numerical computations are done in FORTRAN. 

We have implemented in the framework of the {\tt SANC} system processes such  as
\begin{eqnarray}
\gamma(p_1,\lambda_1) + \gamma(p_2,\lambda_2) \longrightarrow Z(p_3,\lambda_3)
+ Z(p_4,\lambda_4)
\label{ggZZ}
\end{eqnarray}
($\lambda_i,\ (i=1,2,3,4)$ are the helicities of the external particles)
in the Standard Model (SM) at the one-loop level of accuracy in $R_{\xi}$-gauge 
through fermion loops and corresponding precomputation blocks.

The computations of these processes take into account non-zero masses of
loop-fermions.  
Our previous evaluation for processes 
$\gamma + \gamma \longrightarrow \gamma + \gamma $
and 
$ \gamma + \gamma \longrightarrow \gamma
+ Z, Z \longrightarrow \gamma\gamma\gamma $ 
was presented in~\cite{Bardin:2009gq} and~\cite{Bardin:2012mc}.
%More details are presented in our first publications
%about the four boson sector in section~\ref{HA}~\cite{Bardin:2009gq}. 
The additional precomputation modules used to calculate massive
fermion-box diagrams are briefly described. 
%Also in {\tt SANC} system one has an opportunity of sending the analytical
%results to numerical  
%evolution~\cite{Andonov:2004hi}.

We discuss the covariant and tensor structures 
and present them in a compact form.
The helicity amplitude  approach and their expressions are given.

First of all  we calculate $bbbb \to 0$ in the multi-channel approach and get
the main object form factors (FF) for the annihilation to vacuum.
In the  second step we produce the FF of the real channel \ref{ggZZ} by a
permutation of the arguments $s,t$ and $u$. 

The 4-momenta of incoming bosons are denoted by $p_1$ and $p_2$, of the
outgoing ones by $p_3$ and $p_4$. 
%The amplitudes are calculated for scattering of real photons, that is 
%$p_1^2=0\,,~p_2^2=0\,,~p_3^2=0\,,~p_4^2=0$. 

The 4-momenta conservation reads
\begin{equation}
p_1+p_2-p_3-p_4=0\,. \nonumber
\end{equation}
The Mandelstam variables are\footnote{Note, that in {\tt SANC} we use the Pauli
  metric.} 
\begin{eqnarray}
s=~(p_1+p_2)^2,~ t=-(p_1-p_3)^2, ~ u=-(p_1-p_4)^2. \nonumber
\end{eqnarray}

%There is a rich world literature devoted to these processes.
Whenever possible, we compare the results with those existing in the literature.

To check our numerical and analytical results we compare them with other
independent calculations \cite{Diakonidis:2006cv}.

The paper is organized as follows.
In section \ref{sec:PS} we briefly describe the precomputation strategy (see also,
\cite{Andonov:2004hi}). 
In section~\ref{sec:CA} we discuss the tensor structures of
covariant amplitudes for these processes and present their compact form.
The idea of form factors is described in section~\ref{sec:FFs}.
In section~\ref{sec:HA} we consider the sets of the corresponding helicity amplitudes.
In the conclusions, section \ref{sec:RaC}, we present
the comparison of our results with those  existing in the literature. 

%=========================================================================

\section{Precomputation strategy}
\label{sec:PS}

In this section we briefly describe the modules relevant for $ bb \to bb $,
($b$ - is $A$ or $Z$).
The contributions of fermionic loop boxes form a gauge-invariant and UV-finite
subset.  

In {\tt SANC}, the idea of precomputation becomes vitally
important for boxes~\cite{Andonov:2004hi}. 
The calculation of some boxes for some particular processes takes so much time
that an external user should  
refrain from repeating the precomputation. Furthermore, the richness of boxes
requires a classification.  
Depending on the type of external lines ($f$ -- fermion or $b$ -- boson), we distinguish
three large classes  
of boxes: $ffff$, $ffbb$ and $bbbb$.

The precomputation file {\tt bbbb Box}, i.e. {\tt AAAA Box, AAAZ Box, AAZZ
  Box} contains the sequence 
of procedures to calculate the covariant amplitude. 
At this step we suppose that all momenta are incoming (denoted by $p_i$) and
photons are not  on-mass-shell.
Therefore, these results can be used for other processes which need these
parts as  building blocks.

When we implement the processes $ bb \to bb$, we use this building block
several times  
by replacing the incoming momenta $p_i$ by the corresponding kinematical
momenta $k_i$, 
calculate FFs by the module {\tt bb->bb (FF)}, then helicity amplitudes by
the modules {\tt bb->bb (HA)}
and finally the differential and total process cross section by the module
{\tt bb->bb (XS)}. 

%=========================================================================

\section{Covariant amplitude}
\label{sec:CA}

The covariant one-loop amplitude (CA) corresponds to the result of the
straightforward standard calculation  
of all diagrams contributing to a given process at the %tree (Born) and 
one-loop (1-loop) level. 
The CA is being represented in a certain basis, made of strings of Dirac
matrices and/or  
external momenta (structures), contracted with polarization vectors of vector
bosons, $\epsilon(k)$, if any.  

A CA can be written in an explicit form with the aid of scalar FFs. 
\begin{eqnarray}
{\cal A}_{\gamma \gamma \to ZZ} = 
4e^4Q^4_f\sum\limits_{i=1}^{20}
\left[{\cal F}^{\tt bos}_{i}\left(s\,,t\,,u\right)+{\cal F}^{\tt
    fer}_{i}\left(s\,,t\,,u\right)\right]  
          T_{i}^{\alpha\beta\mu\nu}\,.\nonumber
\end{eqnarray}

All masses, kinematical factors and coupling constant and other parameter
dependences are included into these FFs ${\cal F}_{i}$, but tensor structure
with Lorenz indexes made of strings of Dirac matrices are given by basis.
The number of form factors is equal to the number of independent structures.

The covariant amplitude for the channel 
$ b b \rightarrow b b$, i.e. $ \gamma \gamma \rightarrow ZZ $ can be obtained 
from annihilation to the vacuum with
the following permutations of the 4-momenta:
\begin{eqnarray}
&& p_1 \to p_1, ~
 p_2 \to  p_2, ~ p_3 \to   -  p_3, ~
 p_4 \to          -  p_4.    \nonumber
\end{eqnarray}
 
 To describe the tensor structures of the CA for {$AA \to ZZ$} 
we introduce the following five auxiliary tensorial strings:
\begin{eqnarray}
\tau_1^{ij} &=& p^{1i}p^{2j} + \frac{1}{2} s \delta^{ij}, 
\quad
\tau_2^{ij}  = p^{2i} p^{3j} - \frac{1}{2}(\mz^2-t)\delta^{ij}, 
\nonumber \\
\tau_3^{ij}  &=& p^{1i} p^{3j} - \frac{1}{2}(\mz^2-u)\delta^{ij}, 
\nonumber \\
\tau_4^{i}  &=& p^{3i} +\frac{\mz^2-t}{s} p^{1i}, \quad
\tau_5^{i}  =  p^{3i}+\frac{\mz^2-u}{s} p^{2i}.   
\nonumber
\end{eqnarray}

Our basis of CA is given by:
\begin{eqnarray}
T_1^{\alpha\beta\mu\nu} &=&  
 \tau_1^{\beta\alpha} p^{1\mu}p^{1\nu},\quad
T_2^{\alpha\beta\mu\nu} =
\tau_1^{\beta\nu} \tau_1^{\mu\alpha},\quad
%\nonumber \\
T_3^{\alpha\beta\mu\nu} =
\tau_1^{\beta\nu} \tau_3^{\mu\alpha},
\quad
T_4^{\alpha\beta\mu\nu} =
\tau_1^{\beta\mu} \tau_1^{\nu\alpha}, 
\nonumber \\
T_5^{\alpha\beta\mu\nu} &=&
\tau_1^{\beta\mu} \tau_3^{\nu\alpha},\quad
T_6^{\alpha\beta\mu\nu}  =
\tau_1^{\beta\alpha} p^{2\mu} p^{2\nu}, 
%\nonumber \\
\quad
T_7^{\alpha\beta\mu\nu} =
\frac{1}{2}~p^{2\mu}
\left(2 p^{2\nu}\tau_3^{\beta\alpha}+s\delta^{\beta\nu}\tau_5^{\alpha}\right),
\nonumber \\
T_8^{\alpha\beta\mu\nu} &=&
\frac{1}{2}~p^{1\mu}
\left(2 p^{1\nu}\tau_2^{\alpha\beta}+s\delta^{\alpha\nu}\tau_4^{\beta} \right),
%\nonumber \\
\quad
T_9^{\alpha\beta\mu\nu} =
\tau_3^{\nu\alpha} 
            p^{1\mu}
\tau_4^{\beta},
\quad
T_{10}^{\alpha\beta\mu\nu} =
\tau_1^{\mu\alpha} \tau_2^{\nu\beta},
\nonumber \\
T_{11}^{\alpha\beta\mu\nu} &=&
\tau_2^{\nu\beta}\tau_3^{\mu\alpha},  
\quad
T_{12}^{\alpha\beta\mu\nu} =
\tau_1^{\nu\alpha} \tau_2^{\mu\beta},
%\nonumber \\
\quad
T_{13}^{\alpha\beta\mu\nu} =
\tau_2^{\mu\beta} \tau_3^{\nu\alpha}, 
\quad
T_{14}^{\alpha\beta\mu\nu} =
\tau_2^{\nu\beta} 
     p^{2\mu}\tau_5^{\alpha},
\nonumber \\
T_{15}^{\alpha\beta\mu\nu} &=&
  \delta^{\alpha\beta}p^{1\mu}p^{2\nu}-\delta^{\alpha\mu}p^{1\beta}p^{2\nu}
- \delta^{\beta\nu}\tau_1^{\mu\alpha},
\quad
T_{17}^{\alpha\beta\mu\nu} =
      \left( \delta^{\alpha\mu} p^{1\nu} - \delta^{\alpha\nu} p^{1\mu}\right) 
\tau_4^{\beta},
\nonumber \\
T_{16}^{\alpha\beta\mu\nu} &=&
   \delta^{\alpha\beta} p^{1\nu} p^{2\mu} 
 - \delta^{\alpha\nu} p^{1\beta} p^{2\mu}
 - \delta^{\beta\mu} \tau_1^{\nu\alpha},
%\nonumber \\
\quad
T_{18}^{\alpha\beta\mu\nu} =
     \left(\delta^{\beta\mu} p^{2\nu} - \delta^{\beta\nu} p^{2\mu}\right)
\tau_5^{\alpha},
\nonumber \\
T_{19}^{\alpha\beta\mu\nu} &=&
 \tau_1^{\beta\alpha}    \delta^{\mu\nu},
%\nonumber \\
\quad
T_{20}^{\alpha\beta\mu\nu} =
\delta^{\mu\nu}
\left(\frac{\mz-t}{s}\tau_3^{\alpha\beta}+\tau_5^{\alpha} p^{3\beta}\right).
\nonumber
\end{eqnarray}

%=========================================================================

\section{Form factors}
\label{sec:FFs}

In the multi-channel approach  we calculate $bbbb \to 0$, and get
the main object formfactors (FF) for the annihilation into the vacuum.
The form factors ${\cal F}_{i}$ are the scalar coefficients in front of basis
structures of the CA.  
They are presented as combinations of scalar PV functions $A_0$, $B_0$,
$C_0$, $D_0$~\cite{Passarino:1978jh},  
and depend on invariants $s\,,t\,,u\,,$ and also on fermion and boson
masses. They do not contain  ultraviolet poles.
These one-loop scalar form factors can be used for any cross channel
after an appropriate 
permutation of their arguments $s,t,u$.

Explicit expressions for the boson and fermion parts of the form factors are
not shown in this article because they are very cumbersome.
A complete answer for ${\cal F}_{i}$ can be found in the package which is
downloadable from the homepage of the computer system {\tt SANC}.
For massless loop fermions the FFs are rather compact for $\gamma\gamma \to
\gamma\gamma$,  
see~\cite{Bardin:2009gq}.
Note that the expression for the amplitude of boson diagrams are similar to
those for the fermion diagrams except for the explicit representation of form factors.

%=========================================================================

\section{Helicity amplitudes}
\label{sec:HA}

In {\tt SANC} we use the helicity amplitude approach.
In the expression for CA, as one can see in subsection~\ref{sec:CA}, one has tensor
structures and a set of scalar FFs.
To calculate an observable quantity, such as cross section, one needs to take
the square of the amplitude,  
calculate products of Dirac spinors and contract Lorenz indices with
polarization vectors.  In the standard approach of taking amplitude square 
one gets squares for each diagram and their interferences. This leads to a
large number of terms. 

In the helicity amplitude approach we also derive tensor structures and
FFs. But the next step  
is a projection to the helicity basis and as a result one gets a set of
non-interfering amplitudes, since 
all of them are characterized by different sets of helicity quantum numbers.
In this approach we can separate the calculations of Dirac spinors and the
contractions of Lorenz indices  
from calculations of FFs. We can do this before taking the amplitude squares.
So, proceeding in this way, we get a profit on calculation time (a smaller number
of terms due to zero interference) and also a clearer step-by-step control.

In the {\tt SANC} system helicity amplitudes are the result of an application 
of the procedure {\tt TRACEHelicity.prc}. A description of the main {\tt SANC}
procedures is given in \cite{Bardin:2009gq}.

In this section we collect the analytical expressions of the HAs.
The total number of the HA is 36.

 We have verified the analytical zero between the cross sections of
 \cite{Diakonidis:2006cv} and {\tt SANC}. 
\begin{eqnarray}
\gamma(p_1,\lambda_1) + \gamma(p_2,\lambda_2) \longrightarrow Z(p_3,\lambda_3)
+ Z(p_4,\lambda_4)  \nonumber
\end{eqnarray}
$\lambda_i,\ (i=1,2,3,4)$ are the helicities of the external particles.

The relationship between HA obey parity and Bose symmetries:
\begin{eqnarray}
{\cal H}_{\lambda_1\lambda_2\lambda_3\lambda_4}(s,t,u,\lambda) = {\cal
  H}_{-\lambda_1-\lambda_2-\lambda_3\-\lambda_4}(s,t,u,\lambda),  \nonumber\\
{\cal H}_{\lambda_1\lambda_2\lambda_3\lambda_4}(s,t,u,\lambda) = {\cal
  H}_{\lambda_2\lambda_1\lambda_4\lambda_3}(s,t,u,\lambda). \nonumber
\end{eqnarray}
%\begin{eqnarray}
%{\cal H}_{\lambda_1\lambda_2\lambda_3\lambda_4}(s,t,u,\lambda)&=&{\cal
%  H}_{-\lambda_1-\lambda_2-\lambda_3\-\lambda_4}(s,t,u,\lambda). 
%\nonumber \\
%{\cal H}_{\lambda_1\lambda_2\lambda_3\lambda_4}(s,t,u,\lambda)&=&{\cal
%  H}_{\lambda_2\lambda_1\lambda_4\lambda_3}(s,t,u,\lambda). 

Finally  eight independent HA remain due to parity transformation and
corresponding rotation about the y-axis: 
$${\cal H}_{\substack{{+\pm-\mp}}}(s,t,u,\lambda)
= \phantom{-} {\cal H}_{\substack{{+\pm+\pm}}}(s,t,u,-\lambda)\\
{\cal H}_{\substack{{{+\pm- 0}}}}(s,t,u,\lambda)
=          - {\cal H}_{\substack{{+\pm+0}}}(s,t,u,-\lambda)$$

%\begin{eqnarray}
%{\cal H}_{++--}(s,t,u,\lambda)&=& \phantom{-} {\cal H}_{++++}(s,t,u,-\lambda),
%\nonumber \\
%{\cal H}_{+--+}(s,t,u,\lambda)&=& \phantom{-} {\cal H}_{+-+-}(s,t,u,-\lambda),
%\nonumber \\
%{\cal H}_{++-0}(s,t,u,\lambda)&=&- {\cal H}_{+++0}(s,t,u,-\lambda),
%\nonumber \\
%{\cal H}_{+--0}(s,t,u,\lambda)&=&- {\cal H}_{+-+0}(s,t,u,-\lambda).
%\end{eqnarray}

There are 10 sets of HA for this process. 
Inside each set the HA are equal to each other or replace the sign of $\lambda$, or
 change the sign of $\cos\vartheta_z$:
$${\cal H}_{\substack{{++\pm\pm}\\{--\pm\pm}}}, 
{\cal H}_{\substack{{++\pm\mp}\\{--\pm\mp}}},      
{\cal H}_{\substack{{+-\pm\pm}\\{-+\pm\pm}}}, 
{\cal H}_{\substack{{+-\pm\mp}\\{-+\pm\mp}}}, 
{\cal H}_{\substack{{++\pm 0}\\{--\pm 0}}},$$  
$${\cal H}_{\substack{{++0\pm}\\{--0\pm}}}, 
{\cal H}_{\substack{{\mp\pm 00}}},\\{\cal H}_{\substack{{\mp\mp00}}},
{\cal H}_{\substack{{+-0\pm}\\{-+0\pm}}}, 
{\cal H}_{\substack{{+-\pm 0}\\{-+\pm 0}}}.
$$

The  fully massive case of the analytical expressions of the helicity amplitudes
${\cal H}_{\lambda_1 \lambda_2 \lambda_3 \lambda_4}$ has the following form:
\begin{eqnarray}
{\cal H}_{++++}  &=&
\frac{1}{32} s
~\Biggl\{  
         8                  \left(\cmi {\cal F}_1+\cpl {\cal F}_2\right)
       + 4 \cmi \cpl \frac{\sqrtLsmm}{s} \left({\cal F}_3+{\cal F}_4\right)
       + 16 ( {\cal F}_5 - \frac{\mz^2}{s} {\cal F}_6)
\nll &&
       + 
%\Bigl(
   \cmi \cpl \left[ 2 s ({\cal F}_7-{\cal F}_8-{\cal F}_9+{\cal F}_{10})
-\Lll ({\cal F}_{11}-{\cal F}_{15}+{\cal F}_{16})
+\Ll {\cal F}_{14}
\right]
\nll &&
              -(\cmi)^2 \Ll {\cal F}_{12}
              -(\cpl)^2 \Lll {\cal F}_{13}
%    \Bigr)
\nll &&
       + \frac{\Lll}{2 s} \Bigl[
     (\cmi)^2 \left(   \cpl \sqrtLsmm  {\cal F}_{17}+\Ll  {\cal F}_{18}\right)
  +  (\cpl)^2 \left(\Ll {\cal F}_{19}+ \cmi \sqrtLsmm {\cal F}_{20}\right)
                  \Bigr]  
\Biggr\},
\nonumber \\
{\cal H}_{+++-} &=& 
% \frac{1}{64}~s~\cmi\cpl\Biggl\{ 
 \frac{1}{32}~s~\cmi\cpl\Biggl\{ 
      2 s \left(-{\cal F}_7+ {\cal F}_8+ {\cal F}_9- {\cal F}_{10}\right)
\nll &&
     +\Lll \Bigl[
 \left({\cal F}_{11}-{\cal F}_{12}-{\cal F}_{13}-{\cal F}_{14}-{\cal
   F}_{15}+{\cal F}_{16}\right) 
\nll &&
 -\frac{\sqrtLsmm}{2 s} 
 \left(\cmi {\cal F}_{17}+\cpl {\cal F}_{20}\right)
%+\frac{\Lll}{s} \left({\cal F}_{18}+{\cal F}_{19}\right)
 +\frac{\Lll}{2s}\left({\cal F}_{18}+{\cal F}_{19}\right)
           \Bigr]                 
\Biggr\},
\nonumber \\
{\cal H}_{+-++}  &=&
 \frac{1}{64}~\cmi\cpl\Biggl\{ 
    - 8 \frac{\lambda}{s}  {\cal F}_6
    +2 s \Bigl[
          4 \left({\cal F}_1+{\cal F}_2\right)
        - 4 \frac{\sqrtLsmm}{s} \left({\cal F}_3-{\cal F}_4\right)
\nll &&
         -\Lll {\cal F}_{12}
         -\Ll {\cal F}_{13}
        - \cpl \sqrtLsmm {\cal F}_{14}  
\nll &&
        + \cmi \sqrtLsmm 
% \left({\cal F}_{11}+{\cal F}_{15}-{\cal F}_{16}-\frac{\Lll}{s} {\cal F}_{17}
%  +\Ll {\cal F}_{20}\right)
 \left({\cal F}_{11}+{\cal F}_{15}-{\cal F}_{16}-\frac{\Lll}{2s} {\cal F}_{17}
  +\frac{\Ll}{2s}{\cal F}_{20}\right)
\Bigr]
        +\Lll^2 {\cal F}_{18} 
        +\Ll^2 {\cal F}_{19} 
\Biggr\},
\nonumber \\
%\end{eqnarray}
%\begin{eqnarray}
{\cal H}_{+-+-} &=& 
   \frac{1}{32}(\cmi)^2 s \Bigl\{
            4 ({\cal F}_{1}+{\cal F}_{2} )   
\nonumber \\   
&& + (\sqrtLsmm\cpl-u+t-s)
\left[{\cal F}_{12}+{\cal F}_{13}+\frac{1}{2}\cpl\frac{\sqrtLsmm}{s}({\cal
    F}_{17}+{\cal F}_{20})\right]         
\nonumber \\ 
&&  + \sqrtLsmm \cpl ( -{\cal F}_{11}-{\cal F}_{14}-{\cal F}_{15}+{\cal
  F}_{16} + {\cal F}_{17}) 
           + 2\mz^2({\cal F}_{18}+{\cal F}_{19}) 
 \Bigr\}
\nonumber \\
{\cal H}_{+-00} &=&
       \frac{1}{64} \frac{\cpl \cmi}{\mz^2} \Biggl\{
     4 s^2 ({\cal F}_1+{\cal F}_2)
    +2 \lambda \left[2({\cal F}_3+{\cal F}_4)
                         +\frac{1}{s^2} \Llll {\cal F}_6\right]
\nll &&
-s \left[
          \sqrtLsmm  
%(s\cpl-\Ll)
\Kl 
({\cal F}_{11}+ {\cal F}_{15})
        + \Llll {\cal F}_{12} 
        + 4 s \mz^2 {\cal F}_{13} 
        + \sqrtLsmm \Kll ({\cal F}_{14}-{\cal F}_{16})
\right]
 \nll &&
  +2 \mz^2 \left[ \sqrtLsmm 
%(s \cpl-\Ll) 
\Kl
{\cal F}_{17}
                 +\Llll ({\cal F}_{18}+{\cal F}_{19})
                 -\sqrtLsmm \Kll {\cal F}_{20}\right]
\Biggr\},
\nonumber\\ 
{\cal H}_{+++0} &=&
 \frac{1}{32} \sin\vartheta_z\frac{\sqrt{s}}{\sqrt{2}~\mz}\Biggl\{ 
    2 \Ll \left[2(-{\cal F}_1+{\cal F}_2)
   + \frac{\sqrtLsmm}{s} \left(\cmi {\cal F}_3-\cpl {\cal F}_4\right)\right]
\nll && 
   +2 s \left[ \Kl \left(-{\cal F}_7+{\cal F}_9 \right)
             +\Kll \left( {\cal F}_8-{\cal F}_{10}\right)\right]
   + \cmi \Llll {\cal F}_{12}
\nll && 
%   -2( \Ll^2-\cmi \Llll) {\cal F}_{14}
   -( \Ll^2-\cmi \Llll) {\cal F}_{14}
 -\Lll \Bigl(
    s \cpl  {\cal F}_{13}
   - \Kl ({\cal F}_{11}-{\cal F}_{15})
   - \Kll {\cal F}_{16}
\nll &&
 +\frac{1}{2 s} \left[
   \sqrtLsmm \left(\cmi\Kl {\cal F}_{17}+\cpl\Kll {\cal F}_{20}\right) 
   +\Llll\left(\cmi  {\cal F}_{18}-\cpl  {\cal F}_{19}\right)
                \right]
       \Bigr)  
\Biggr\},
\nonumber 
\end{eqnarray}
\begin{eqnarray}
{\cal H}_{++00} &=&
 \frac{1}{32}\frac{1}{\mz^2}\Biggl\{
       -2 (s^2 \cpl \cmi+\Kll^2) {\cal F}_1
       -2 (s^2 \cpl \cmi+\Kl^2 ) {\cal F}_2
\nll &&
       -2 \cpl \cmi \lambda ({\cal F}_3+{\cal F}_4)
       -4 \Llll     \left[   {\cal F}_5-\frac{\mz^2}{s} {\cal F}_6\right]
       - s \Bigl[\Kll \Kl {\cal F}_7
\nll &&
       -  \Kll^2 {\cal F}_8 
       -  \Kl^2 {\cal F}_9
%       -2*(2*s^2-cpl*cmi*s^2-k3l)*FF10
%     -2*s*(s^2*cos(thz)^2 -sqrtLsmm^2)*FF10
%  ?????
       +(s^2\cos^2\vartheta_z-\lambda){\cal F}_{10}
\nll &&
       +\frac{\cos\vartheta_z}{2} \Ll\Lll \left[   \Kl 
   \left( {\cal F}_{11}-{\cal F}_{15}\right)
       + \Kll  {\cal F}_{16}\right]
\nll &&
       +\frac{1}{2}\cpl \cmi s  \left[\Llll {\cal F}_{12}
       +\Ll\Lll  {\cal F}_{13}\right]
       -\Kll \left(\frac{1}{2}\cos\vartheta_z \Llll
                            -s \sqrtLsmm\right) {\cal F}_{14}
\nll &&
       - \cpl \cmi \mz^2\left[ 
         \sqrtLsmm \Kl {\cal F}_{17}
       +  \Llll ({\cal F}_{18}+{\cal F}_{19})
       -  \sqrtLsmm \Kll {\cal F}_{20}\right]
\Biggr\},
\nonumber \\
{\cal H}_{+-0+} &=&
     \frac{1}{32} \sin\vartheta_z\frac{\sqrt{s}}{\sqrt{2}~\mz}\cpl\Biggl\{
         4 s \left({\cal F}_1+{\cal F}_2\right)
       - 2 \frac{\sqrtLsmm}{s}\left( \Ll {\cal F}_3+\Lll {\cal F}_4\right)
\nll && 
       + s\left[
          \cmi  \sqrtLsmm {\cal F}_{11}
       -  \Lll  {\cal F}_{12}
       - 4 \mz^2 {\cal F}_{13}
       + \cmi \sqrtLsmm \left({\cal F}_{15}-{\cal F}_{16}\right)\right]
\nll &&
       - \sqrtLsmm \Kll {\cal F}_{14}
      - 2 \mz^2\left[
       \cmi  \sqrtLsmm ({\cal F}_{17}-{\cal F}_{20})
      - \Lll {\cal F}_{18}
      - \Ll {\cal F}_{19}\right]
\Biggr\},
\nonumber\\
%\end{eqnarray}
%\begin{eqnarray}
{\cal H}_{++0+} &=&
 \frac{1}{32} \sin\vartheta_z\frac{\sqrt{s}}{\sqrt{2}~\mz}\Biggl\{
        4  \Ll \left({\cal F}_1-{\cal F}_2\right)
       + \frac{2\sqrtLsmm}{s} \Ll \left(\cpl {\cal F}_3-\cmi {\cal F}_4\right)
\nll &&
       + 2 s \Bigl[
\Kll 
\left({\cal F}_7-{\cal F}_8  \right) 
       -\Kl  
\left({\cal F}_9-{\cal F}_{10}\right)
       -  \frac{\cmi}{2}  \Ll {\cal F}_{12}
\nll &&
       + 2 \cpl  \mz^2 {\cal F}_{13}\Bigr] 
       +  \Kll \Ll {\cal F}_{14} 
       - 2 \mz^2\Bigl[
        2 s  \cos\vartheta_z (  {\cal F}_{11}-{\cal F}_{15}+{\cal F}_{16})
\nll &&
       - \cpl \cmi  \sqrtLsmm \left({\cal F}_{17}-{\cal F}_{20} \right)
       - \Ll \left( \cmi{\cal F}_{18}-\cpl {\cal F}_{19}\right)  
      \Bigr]
\Biggr\},
\nonumber \\
{\cal H}_{+-+0} &=&
 \frac{1}{32} \sin\vartheta_z\frac{\sqrt{s}}{\sqrt{2}~\mz}\cmi\Biggl\{
              4 s \left({\cal F}_1+{\cal F}_2\right)
            - \frac{2 \sqrtLsmm}{s} \left[\Ll {\cal F}_3
                                 -\Lll {\cal F}_4 \right]
 \nll &&
    -  \sqrtLsmm
%(s \cpl-\Ll)
\Kl 
\left({\cal F}_{11}+{\cal F}_{15}\right)      
    -    \Llll {\cal F}_{12}
    -  s \Ll {\cal F}_{13}
    -  \cpl s \sqrtLsmm  {\cal F}_{14}
\nll &&
    +\frac{1}{2 s} \Bigl[  \sqrtLsmm \Kll 
              \left(2s {\cal F}_{16}-\Ll{\cal F}_{20}\right)
    + \sqrtLsmm 
%(s \cpl-\Ll) 
\Kl
\Lll {\cal F}_{17}
    +     \Llll \left(\Lll {\cal F}_{18}
    +    \Ll {\cal F}_{19}\right)
 \Bigr]
\Biggr\},
\nonumber
%\nonumber \\
%\hline
%%------------------ 10
%{\cal H}_{+---}  &=&
% \frac{1}{32}\cmi\cpl \Biggl\{ 
%        4 \left[ s ({\cal F}_1+{\cal F}_2)
%       + \sqrtLsmm ({\cal F}_3-{\cal F}_4)
%       - \frac{\lambda}{s} {\cal F}_6 \right] 
%\nll &&     
%      -s\Bigl[
%        \cpl \sqrtLsmm \left({\cal F}_{11}+{\cal F}_{15}-{\cal F}_{16}\right) 
%      +  \Ll          {\cal F}_{12}
%      +  \Lll         {\cal F}_{13}
%\nll && 
%      - \cmi \sqrtLsmm  {\cal F}_{14}
%       \Bigr]
%      +\frac{1}{2}\left( \cpl \sqrtLsmm\left[\Ll  {\cal F}_{17}-\Lll{\cal F}_{20}\right]
%      +                  \Ll^2  {\cal F}_{18}
%      +                  \Lll^2 {\cal F}_{19}\right)
%\Biggr\}.
%%   [Amp_+1-1+1-1] =\
\end{eqnarray}

where
\begin{eqnarray}
 u &=& \cos\vartheta_z \sqrtLsmm + t, \quad c^{\pm} = 1\pm\cos\vartheta_z
\nonumber \\
 \Ll &=& s-\sqrtLsmm,\quad \Lll =s+\sqrtLsmm,\quad \Llll =s^2+\lambda,
\nll
\Kl &=&
s \cpl-\Ll, \quad
\Kll =
s \cpl-\Lll,\quad
 \lambda = s (s-4 \mz^2).\nonumber
\end{eqnarray}

The definitions of cross-sections $\sigma_{ij}$ were given in \cite{Gounaris:1999hb}: 
\begin{eqnarray}
  \frac{d\sigma_{0}}{d \cos{\vartheta^*}} &=&
  \left( \frac{\beta_{Z}}{64\pi s} \right)
  \sum_{\lambda_3 \lambda_4} 
  \left[ |{\cal H}_{++\lambda_3 \lambda_4}|^2 + |{\cal H}_{+-\lambda_3 \lambda_4}|^2\right],
  \nll
  \frac{d\sigma_{22}}{d \cos{\vartheta^*}} &=&
  \left( \frac{\beta_{Z}}{64\pi s} \right)
  \sum_{\lambda_3 \lambda_4} 
  \left[ |{\cal H}_{++\lambda_3 \lambda_4}|^2 - |{\cal H}_{+-\lambda_3 \lambda_4}|^2\right],
  \nll
  \frac{d\sigma_{33}}{d \cos{\vartheta^*}} &=&
  \left( \frac{\beta_{Z}}{64\pi s} \right)
  \sum_{\lambda_3 \lambda_4} 
  {Re}\left[ {\cal H}_{+-\lambda_3 \lambda_4} {\cal H}^*_{-+\lambda_3 \lambda_4} \right],
  \nll
  \frac{d\sigma_{3}}{d \cos{\vartheta^*}} &=&
  \left( \frac{-\beta_{Z}}{32\pi s} \right)
  \sum_{\lambda_3 \lambda_4} 
  {Re}\left[ {\cal H}_{++\lambda_3 \lambda_4} {\cal H}^*_{-+\lambda_3 \lambda_4} \right],
\nonumber      
\end{eqnarray}

where $$\beta_Z=1 - \frac{M^2_Z}{s}.$$

%=========================================================================
\section{Conclusions}
\label{sec:RaC}

This paper is devoted 
to the description of implementing the complete one-loop electroweak 
calculations for the process  $\gamma\gamma \to ZZ$  into the {\tt SANC} framework. 
We presented analytical expressions for the Covariant  Amplitude 
and for the Helicity Amplitudes.
To be assured of the correctness of our analytical results, we
checked the independence of the form factors on gauge parameters 
(all calculations were done in $R_{\xi}$ gauge),
the validity of Ward identities for covariant amplitudes and, finally, the
{\tt SANC} results for this processes were compared with other independent calculations
\cite{Diakonidis:2006cv},\cite{Gounaris:1999hb} (see figures 1-4).
For all of the contributions good agreement was obtained with the
results, given in the literature.

We begin to develop MC {\tt SANC} generator at the one-loop level
taking into account the polarization
for future linear $e^ +e^-$ colliders -- ILC and CLIC.
This study for Bhabha process we are going to present in the near future.
To fill the generator we have  library for the complete one-loop electroweak
modules.

The full version MC {\tt SANC} generator will contain the processes
$\gamma\gamma \to \gamma\gamma (\gamma Z,ZZ)$ and some part of
the library of the necessary  complete one-loop modules we present here.

\nocite{*}
\bibliographystyle{pepan}
\bibliography{paper}

\newpage
%============================= Fig. 1 ================================
\begin{figure}[!h]
\begin{center}
\includegraphics[width=80mm,height=60mm]{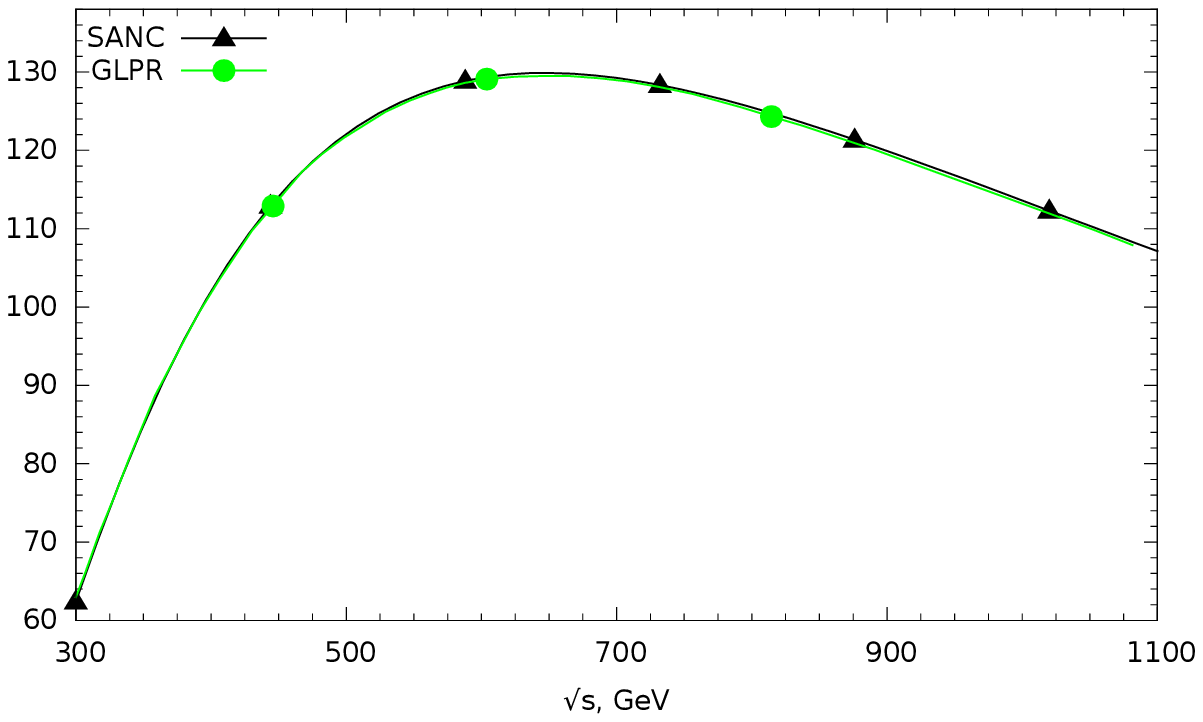}
\vspace{-3mm}
\caption{Contribution to the integrated $\sigma_0$ cross section.
Black line is {\tt SANC} result,
green line is result from paper
~\cite{Gounaris:1999hb}.
}
\end{center}
\labelf{fig01}
\vspace{-5mm}
\end{figure}
%============================= Fig. 1 ================================
%============================= Fig. 2 ================================
\begin{figure}[!h]
\begin{center}
\includegraphics[width=80mm,height=60mm]{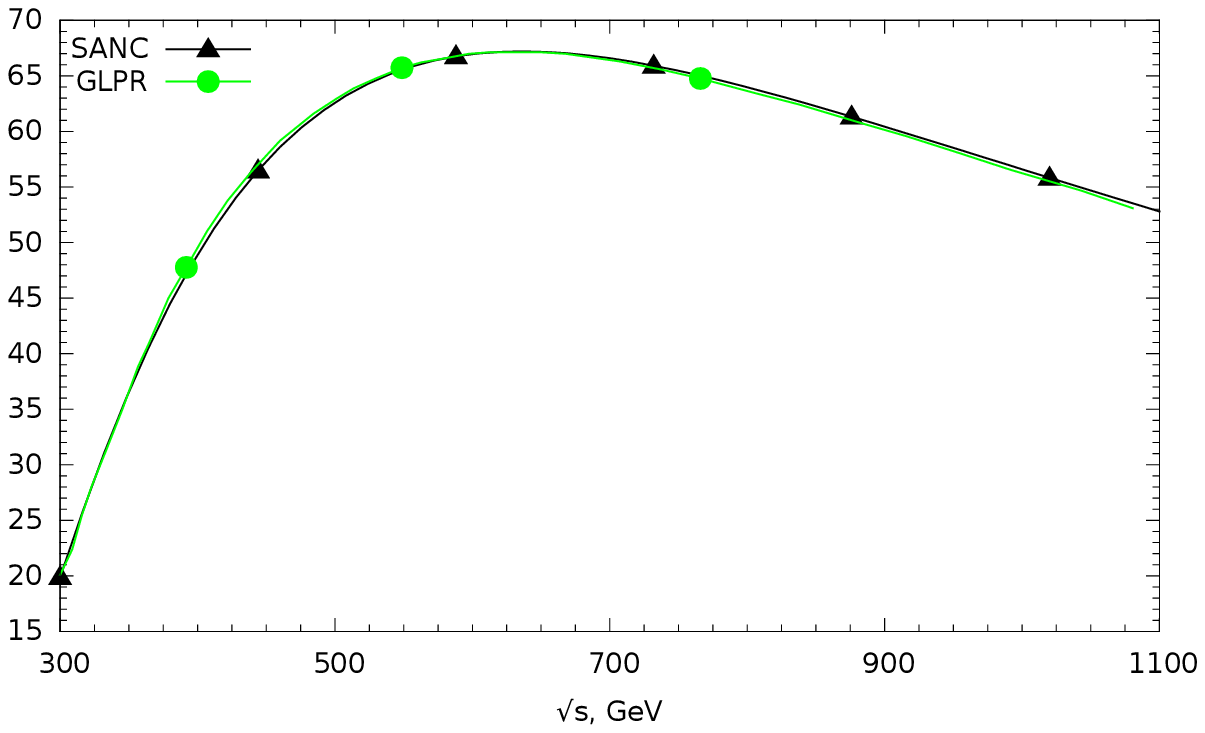}
\vspace{-3mm}
\caption{Contribution to the integrated $\sigma_{22}$ cross section.
Black line is {\tt SANC} result, green line is result from paper
  ~\cite{Gounaris:1999hb}.
}
\end{center}
\labelf{fig02}
\vspace{-5mm}
\end{figure}
%============================= Fig. 2 ================================
%============================= Fig. 3 ================================
\begin{figure}[!h]
\begin{center}
\includegraphics[width=80mm,height=60mm]{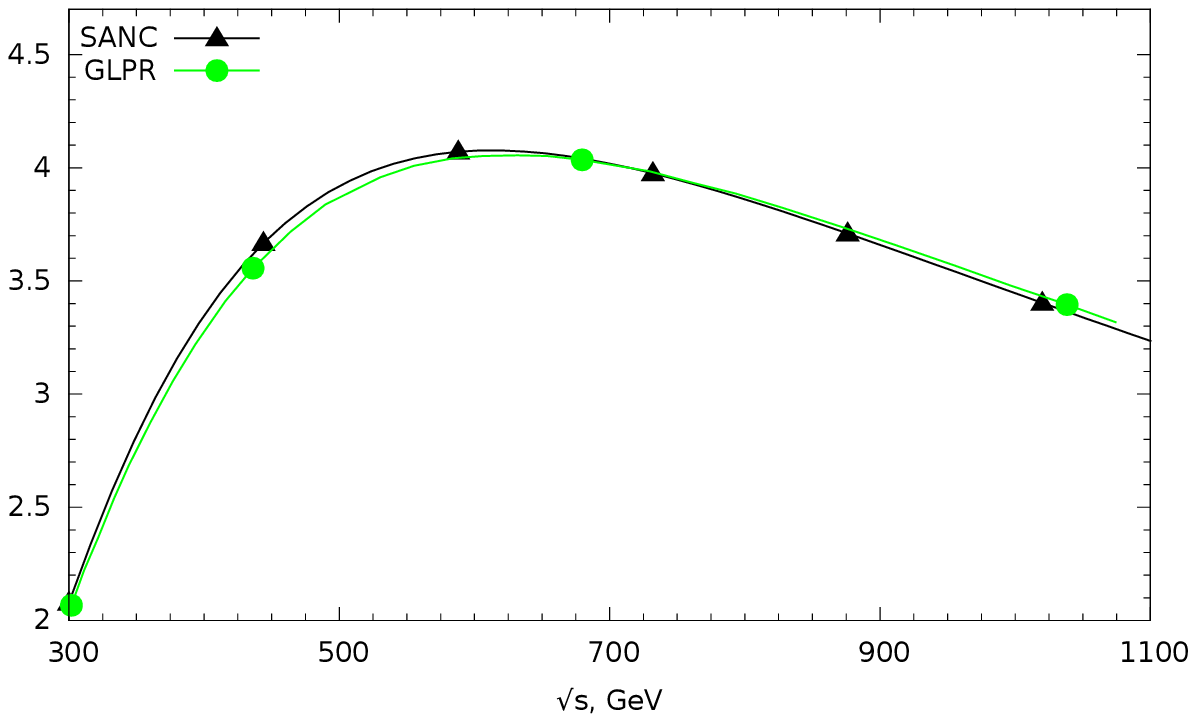}
\vspace{-3mm}
\caption{Contribution to the integrated $\sigma_{33}$ cross section.
Black line is {\tt SANC} result,
green line is result from paper
~\cite{Gounaris:1999hb}.
}
\end{center}
\labelf{fig03}
\vspace{-5mm}
\end{figure}
%============================= Fig. 3 ================================
%============================= Fig. 4 ================================
\begin{figure}[!h]
\begin{center}
\includegraphics[width=80mm,height=60mm]{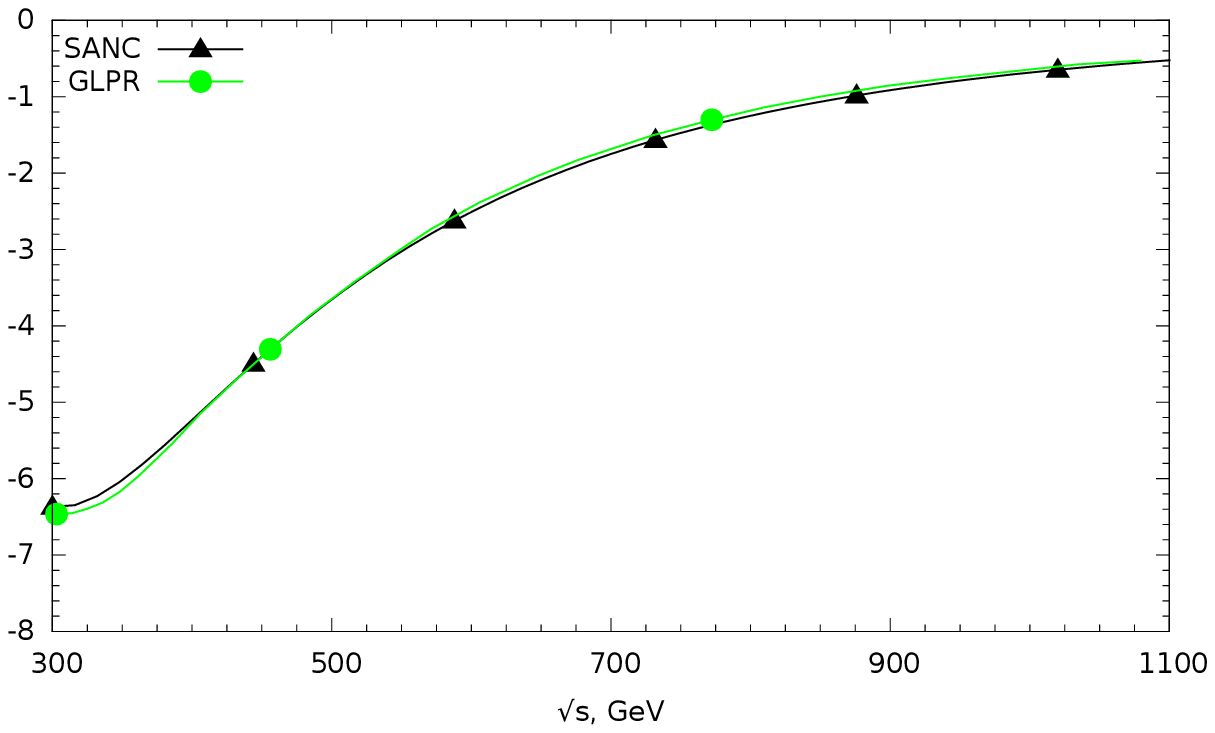}
\vspace{-3mm}
\caption{Contribution to the integrated $\sigma_{3}$ cross section.
Black line is {\tt SANC} result,
green line is result from paper
~\cite{Gounaris:1999hb}.
}
\end{center}
\labelf{fig04}
\vspace{-5mm}
\end{figure}
%============================= Fig. 4 ================================

%=========================================================================

\end{document}